\documentclass[conference]{IEEEtran}
\IEEEoverridecommandlockouts
\usepackage{cite}
\usepackage[justification=centering]{caption}
\usepackage{amsmath,amssymb,amsfonts}
\usepackage{algorithmic}
\usepackage{graphicx}
\usepackage{textcomp}
\usepackage{tabu}
\usepackage{xcolor}
\def\BibTeX{{\rm B\kern-.05em{\sc i\kern-.025em b}\kern-.08em
    T\kern-.1667em\lower.7ex\hbox{E}\kern-.125emX}}

\makeatletter
\newcommand*\titleheader[1]{\gdef\@titleheader{#1}}
\AtBeginDocument{%
	\let\st@red@title\@title%
	\def\@title{%
		\bgroup\normalfont\small\centering\@titleheader\par\egroup
		\vskip0.6em\st@red@title}
}
\makeatother

\title{Securing Fog-to-Things Environment Using Intrusion Detection System Based On Ensemble Learning
{\footnotesize \textsuperscript{}}
\thanks{This research was supported by the NSERC B-CITI CRDPJ 501617-16 grant.}
}
\titleheader{This is the authors’ version of the paper that has been accepted for publication in IEEE Wireless Communications and Networking Conference,  15- 18 April 2019, Marrakesh, Morocco}

\begin{document}
\author{
    \IEEEauthorblockN{
        Poulmanogo Illy\IEEEauthorrefmark{1}, 
        Georges Kaddoum\IEEEauthorrefmark{1}, 
        Christian Miranda Moreira\IEEEauthorrefmark{1},
        Kuljeet Kaur\IEEEauthorrefmark{1},
        and Sahil Garg\IEEEauthorrefmark{1}
        }
    \IEEEauthorblockA{
        \IEEEauthorrefmark{1}
        Electrical Engineering Department, \'{E}cole de Technologie Sup\'{e}rieure, Montr\'{e}al, Canada.
        }
     \IEEEauthorblockA{
        Email: poulmanogo.illy.1@ens.etsmtl.ca
        }
}
\maketitle

\begin{abstract}
The growing interest in the Internet of Things (IoT) applications is associated with an augmented volume of security threats. In this vein, the Intrusion detection systems (IDS) have emerged as a viable solution for the detection and prevention of malicious activities. Unlike the signature-based detection approaches, machine learning-based solutions are a promising means for detecting unknown attacks. However, the machine learning models need to be accurate enough to reduce the number of false alarms. More importantly, they need to be trained and evaluated on realistic datasets such that their efficacy can be validated on real-time deployments. Many solutions proposed in the literature are reported to have high accuracy but are ineffective in real applications due to the non-representativity of the dataset used for training and evaluation of the underlying models. On the other hand, some of the existing solutions overcome these challenges but yield low accuracy which hampers their implementation for commercial tools. These solutions are majorly based on single learners and are therefore directly affected by the intrinsic limitations of each learning algorithm. The novelty of this paper is to use the most realistic dataset available for intrusion detection called NSL-KDD, and combine multiple learners to build ensemble learners that increase the accuracy of the detection. Furthermore, a deployment architecture in a fog-to-things environment that employs two levels of classifications is proposed. In such architecture, the first level performs an anomaly detection which reduces the latency of the classification substantially, while the second level, executes attack classifications, enabling precise prevention measures.  Finally, the experimental results demonstrate the effectiveness of the proposed IDS in comparison with the other state-of-the-arts on the NSL-KDD dataset.
\end{abstract}
\bigskip
\vspace{-0.1 cm}
\section{Introduction}
The Internet of Things (IoT) paradigm offers prodigious opportunities to the industries \cite{perera2015emerging}. This technology is expected to be further active with the imminent Fifth-Generation (5G) mobile communications system \cite{palattella2016internet}. However, the massive deployment of IoT networks and their usage in critical domains such as smart housing, smart transportation, and e-health, results in the generation of abundant sensitive data on real-time basis. Due to this reason, these networks are deemed to be one of the most vulnerable sites for different security attacks and risks.
To tackle this issue, many research studies have been focused on the first security layer, \textit{i.e.}, the prevention layer. Thus, stronger authentication, authorization, and cryptography techniques have been proposed in the literature. However, despite the deployment of such strong security measures, a system can still be compromised by an enduring adversary using advanced techniques or high computational resources. Therefore, under any prevention layer, there must be an intrusion detection layer. This is the motivation for the development of intrusion detection systems (IDS). Majority of intrusion detection solutions deployed commercially implement signature-based approaches. Unlike the signature-based IDS, the machine learning-based IDS are capable of detecting even unknown attacks. Nevertheless, the fundamental challenge in this direction involves the designing of an efficient machine learning based IDS that performs well on real-time data.

The majority of machine learning-based IDS proposed in the literature have been built on KDDCUP\textquotesingle99 dataset \cite{cup1999dataset}. The corresponding evaluations results indicate impressive performances in terms of high accuracy (99\%) and negligible false positive rate (1\%) \cite{shyu2003novel, kim2005genetic, kumar2007network}. Despite their good performances, the existing solutions are still not employed widely in commercial tools, relatively to the signature-based approaches. To understand this situation, the work in \cite{tavallaee2009detailed} conducted a statistical analysis on KDDCUP\textquotesingle99 dataset and found some important issues, mainly induced by a huge number of redundant records. To address these problems, the authors provided a new dataset named NSL-KDD (comprising of KDDTrain+, KDDTest+, and KDDTest-21) that is more realistic and challenging enough to compare different solutions. Based on these refinements, many machine learning methods have been proposed and compared in the literature. In \cite{tavallaee2009detailed}, the authors implemented five different methods, namely  Naive Bayes/Decision-Tree, Random Tree, Decision Tree J48, Random Forest, and Multi-Layer Perceptron on the refined datasets that led to overall accuracy of 82.02\% on KDDTest+ and 66.16\% on KDDTest-21 datasets respectively. To improve the detection, the work in \cite{bajaj2013improving} employed different feature selection metrics at the pre-processing phase for dimensionality reduction on NSL-KDD dataset. Overall, accuracy of 82.32\% and 66.77\% were achieved on KDDTest+ and KDDTest-21 respectively, which is quite a small performance improvement. Ibrahim \textit{et al.} in \cite{ibrahim2013comparison} employed a Self-Organization  Map (SOM) Artificial Neural Network (ANN) which is an unsupervised learning neural network used for classifying the system's input data into normal and abnormal/intrusive instances. In this work, the authors performed only a binary classification, and the model resulted to merely 75.49\% detection rate on KDDTest+, after training it rigorously for an hour. Ingre \textit{et al.} proposed in \cite{ingre2015performance} a deep learning method using a tansig transfer function, Levenberg-Marquardt (LM) and BFGS quasi-Newton Backpropagation algorithm for updating weights and biases. Using the same dataset, the multi-class classification gave 81.2\% of accuracy. Additionally, the authors also performed a binary classification that led to 79.9\% accuracy. 
In another work \cite{yin2017deep}, Yin \textit{et al.} used ANN and applied a Recurrent Neural Network (RNN) after a data pre-processing (normalization) phase. The authors built a multi-class classification and a binary classification models. These models were trained on KDDTrain+ with different numbers of hidden nodes, and different values of learning rates. However, the maximum accuracy achieved was 83.28\%,  for the binary classification on KDDTest+.

\smallskip
\noindent\textbf{Motivation:} To summarize, most of these works were founded on two conventional approaches. Either they tried different learning algorithms and chose the one that performed the best on the test dataset, or they focused on tuning the hyperparameters of one model until they got satisfactory accuracy. However, there are machine learning problems in which even the best learner is not accurate enough. Therefore, many solutions have investigated the combination of learners in different application domains such as image processing, financial forecasting, and weather forecasting \cite{alimoglu1997combining, yu2008forecasting, karvelis2017ensemble}. These ensemble learners have resulted to important improvements over the single learners. The work in \cite{paulauskas2017analysis} analyzed the data pre-processing influence on intrusion detection using NSL-KDD dataset. Among the machine learning methods employed to build and evaluate models on different pre-processed data, ensemble methods were used. The best performance recorded, 84.84\% accuracy for binary classification, was produced by an ensemble model. The base learners in this ensemble complemented each other and reached a higher accuracy. Thus, motivated by this result, the proposed work considers that investigating more diverse base learners, and suitably combining them, will result to ensemble models that perform better on NSL-KDD dataset.

\smallskip
\noindent\textbf{Contributions:} The contributions of this work are three fold: 
\begin{itemize}
\item Firstly, the proposed IDS employs diverse base learners using different algorithms and implements different ensemble methods. On NSL-KDD dataset, the ensemble learners attain higher accuracy than the single learners.
\item Secondly, a deployment architecture in a fog-to-things environment is proposed. It reduces considerably the latency of the IDS by performing anomaly detection at a first stage. In a second stage, it provides attack classification that helps to determine precise prevention measures.
\item  Thirdly, analysis and comparison of the proposed IDS with different solutions in the literature using NSL-KDD dataset has been carried out.
\end{itemize}

\smallskip
\noindent\textbf{Organization:} The rest of the paper is organized as follows. Section II describes the combination of multiple learners to build ensemble learners. Section III presents the proposed IDS solution in fog-to-things environment. A detailed description of the NSL-KDD dataset is done in Section IV. Following this, Section V presents the experimental evaluations of the proposed models and some comparisons with the current state-of-the-art. Finally, Section VI concludes the work.
 \vspace{-0.2 cm}
\section{ combination of Multiple learners }
In this section, the concept of combining multiple learners for achieving higher accuracy is presented.

The machine learning methods use a set of assumptions over the data distribution in order to build the models. In the classification task, sometimes the problem to solve has patterns that differ from the assumptions made in every single machine learning method. Thus, each method is capable to learn only some parts of these patterns and are unable to achieve satisfactory accuracy alone. In this case, by suitably combining different learners, they can complement each other and attain higher accuracy. This idea is reinforced by the ``\textit{No Free Lunch Theorem}'' which advocates the following: ``There is no single learning algorithm that in any domain always induces the most accurate learner''.
However, in order to build an ensemble model that improves the classification accuracy, two essential conditions must be met \cite{alpaydin2009introduction}.

The first condition is to have base learners that are reasonably accurate in their domain of expertise and differ in their decisions. This condition can be achieved by using different algorithms, hyperparameters, input representations, or  training sets. 

\begin{figure}[htbp]
\begin{center}
\includegraphics[width=0.9\columnwidth]{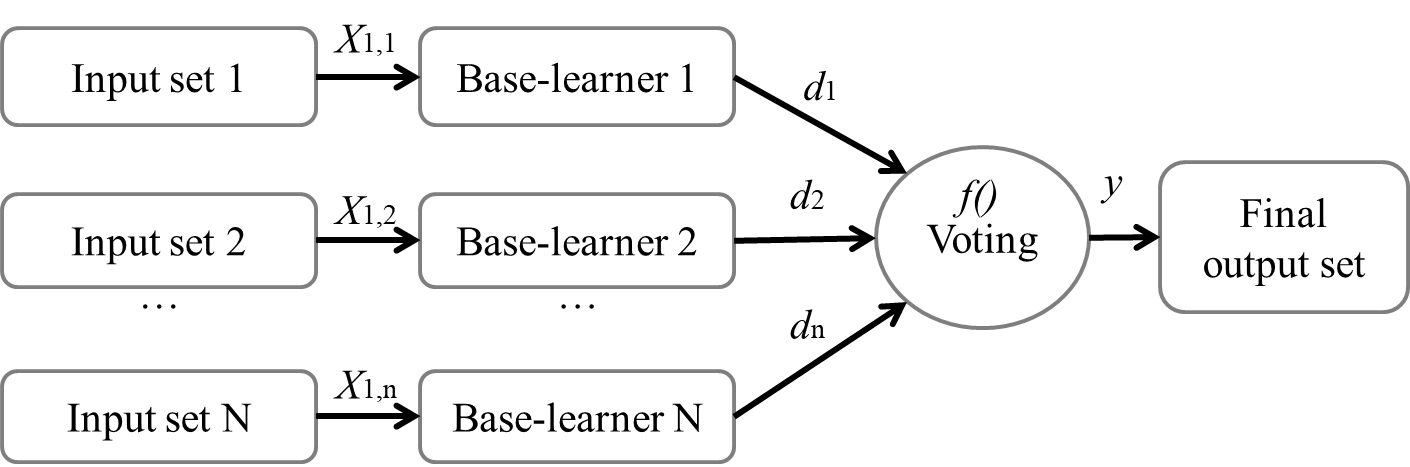}
\end{center}
\small
The inputs observed by learner \(i\) is represented by \(X_1,_i\). The output \(d_i\) is the decision produced by base learner \(i\). The final decision \(y = f(d_1, d_2, ..., d_n |\phi)\), where \(f(.)\) is the combining function with \(\phi\) donating its parameters. This illustration is for the case of a single output. When there are \(K\) outputs, for each learner there are \(d_{ij} (X_1,_i)\), \(i = 1, ..., N\), \(j = 1,...,K\), and combining them, we also generate \(K\) values, \(y_j, j= 1,...,K\). The maximum \(y_j\) is chosen as the predicted class (Choose \(C_j\) if \(y_j = \max_{k=1}^K y_k\)).
\caption{Multiexpert combination scheme.}
\label{Multiexpert combination scheme.}
\end{figure}

\begin{figure}[htbp]
\centering{\includegraphics[width=0.9\columnwidth]{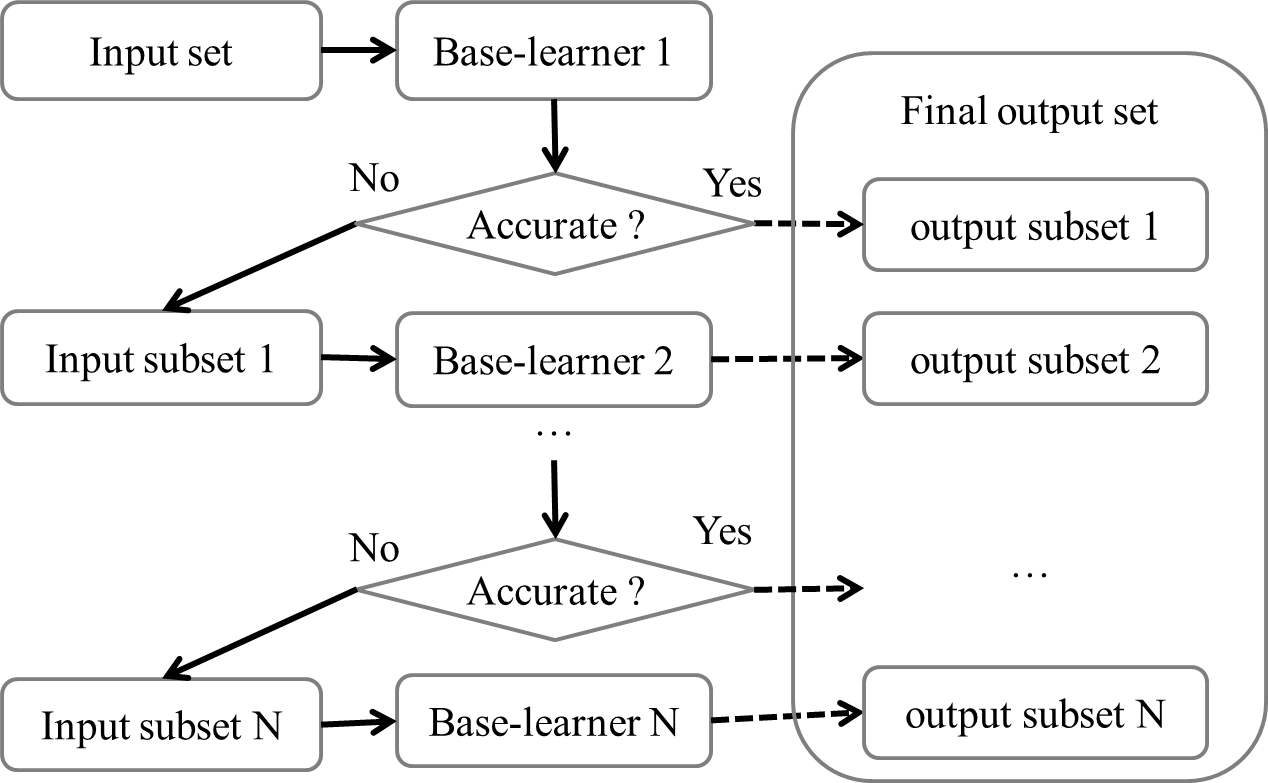}}
\caption{Multistage combination scheme.}
\label{Multistage combination scheme}
\end{figure}

The second condition is how to combine the outputs of multiple base learners to generate the final result. There are two main combination methods used, namely multiexpert and multistage combinations. In the multiexpert combination, the base learners can work in parallel, and all outputs (in the global approach) or selected outputs (in local approach) are then used to generate the final result. Examples of implementation of this combination method are voting and bagging methods. Fig.~\ref{Multiexpert combination scheme.} shows the scheme of this combination method. 
Usually, the operation performed on the outputs of the base learners is a simple function such as sum, weighted sum, median, minimum, maximum or product. Table \ref{Classifier combination rules} defines these functions.
The multistage combination uses a serial approach where each supplementary learner works on the limitations of the previous learners. They are trained or tested only on the instances where the previous base learners were not capable to reach a satisfactory accuracy. An example of implementation of this combination is the boosting method. Fig.~\ref{Multistage combination scheme} illustrates the multistage combination.

\begin{table}[htbp]
\caption{Classifier combination rules.}
\begin{center}
{\tabulinesep=1.2mm
\begin{tabu}{ |l|l| } 
 \hline
 Rule & Fusion function \(f(.)\)  \\ 
 \hline
 Sum & \(y_i = \frac{1}{N}\sum_{j=1}^Nd_{ij}\)   \\ 
 Weighted sum & \(y_i = \sum_j w_jd_{ij}, w_j\geq 0, \sum_j w_j = 1 \)  \\
 Median & \(y_i = \)median\(_jd_{ij}\)\\
 Minimum & \(y_i = \)min\(_jd_{ij}\) \\
  Maximum & \(y_ i = \)max\(_jd_{ij}\) \\
  Product & \(y_i = \prod_jd_{ij}\) \\
 \hline
\end{tabu}}
\end{center}
\label{Classifier combination rules}
\small
In these definitions, \(d_{ij}\) represents the output produced by base learner \(j\) for the instance \(i\). \(N\) is the number of base learners used in the Ensemble. For the Weighted sum, \(w_i\) is the value to weight \(d_{ij}\). \(y_i\) represents the output of the Ensemble leaner for the instance \(i\).
\end{table}
This work used these combination methods to build more accurate classifiers on the NSL-KDD dataset. 
\vspace{-0.2 cm}
\section{Proposed intrusion detection system}
\subsection{The Proposed Method}
\begin{figure}[htbp]
\centering{\includegraphics[width=1\columnwidth]{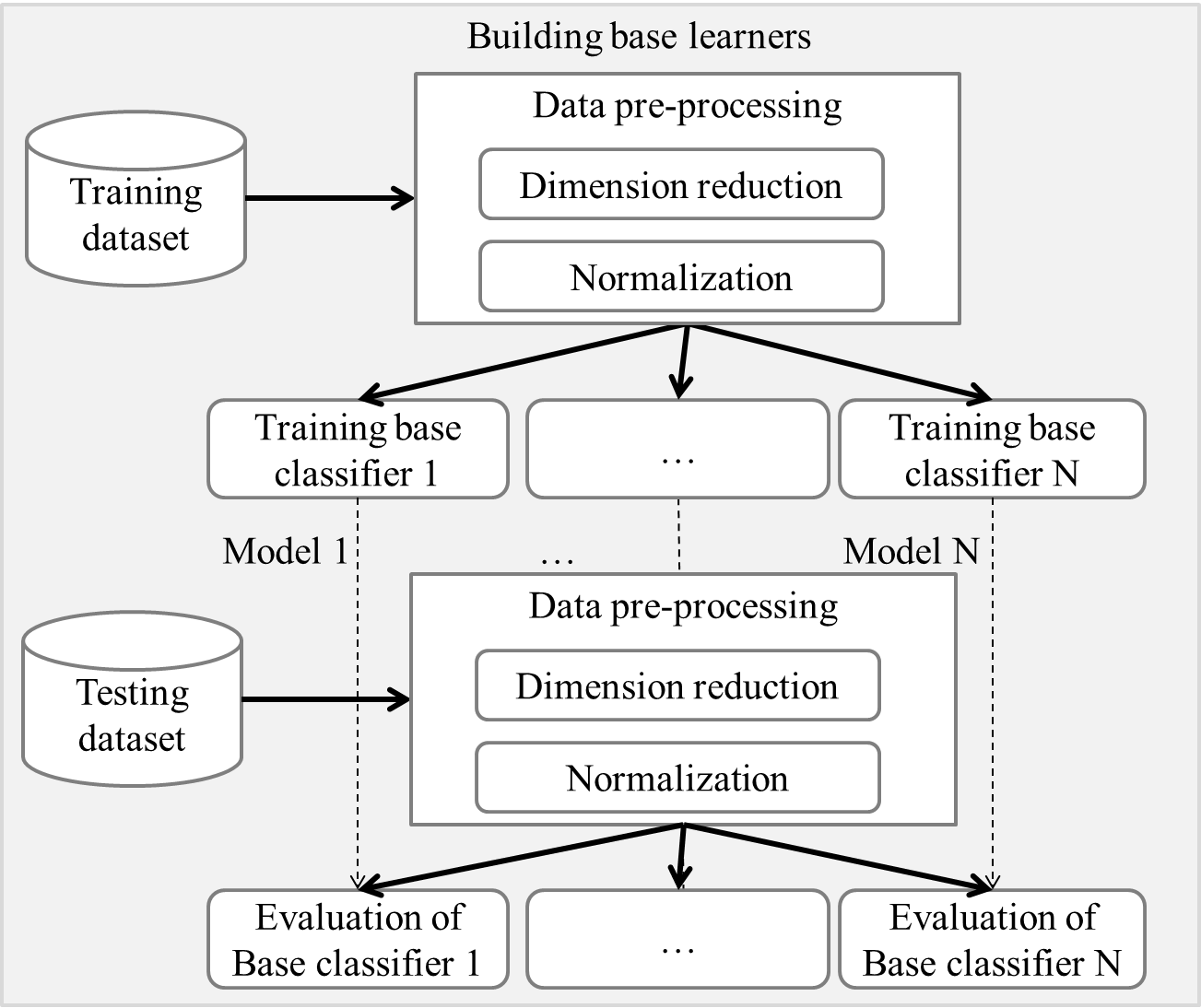}}
\caption{Training and evaluation of base classifiers.}
\label{Building Efficient base classifiers}
\end{figure}

Different steps have been involved in the conception of our ensemble classifiers. At first, we investigated data pre-processing methods including dimensionality reduction and data normalization. Then, diverse base learners using different algorithms have been implemented. We used a Decision Tree, a parametric, a non-parametric, and a deep neural network methods. For each base learner, we tuned the hyperparameters such as the maximum depth for the decision tree algorithm or the number of hidden nodes for the deep learning algorithm. We did not apply advanced tuning methods as we wanted them to be just reasonably accurate, because they are not combined according to their individual accuracy, but for their diversity. After that, the base learners were evaluated in order to analyze their difference in the prediction of each class. Fig.~\ref{Building Efficient base classifiers} describes the procedure for generating the base classifiers.

\begin{figure}[htbp]
\centering{\includegraphics[width=1\columnwidth]{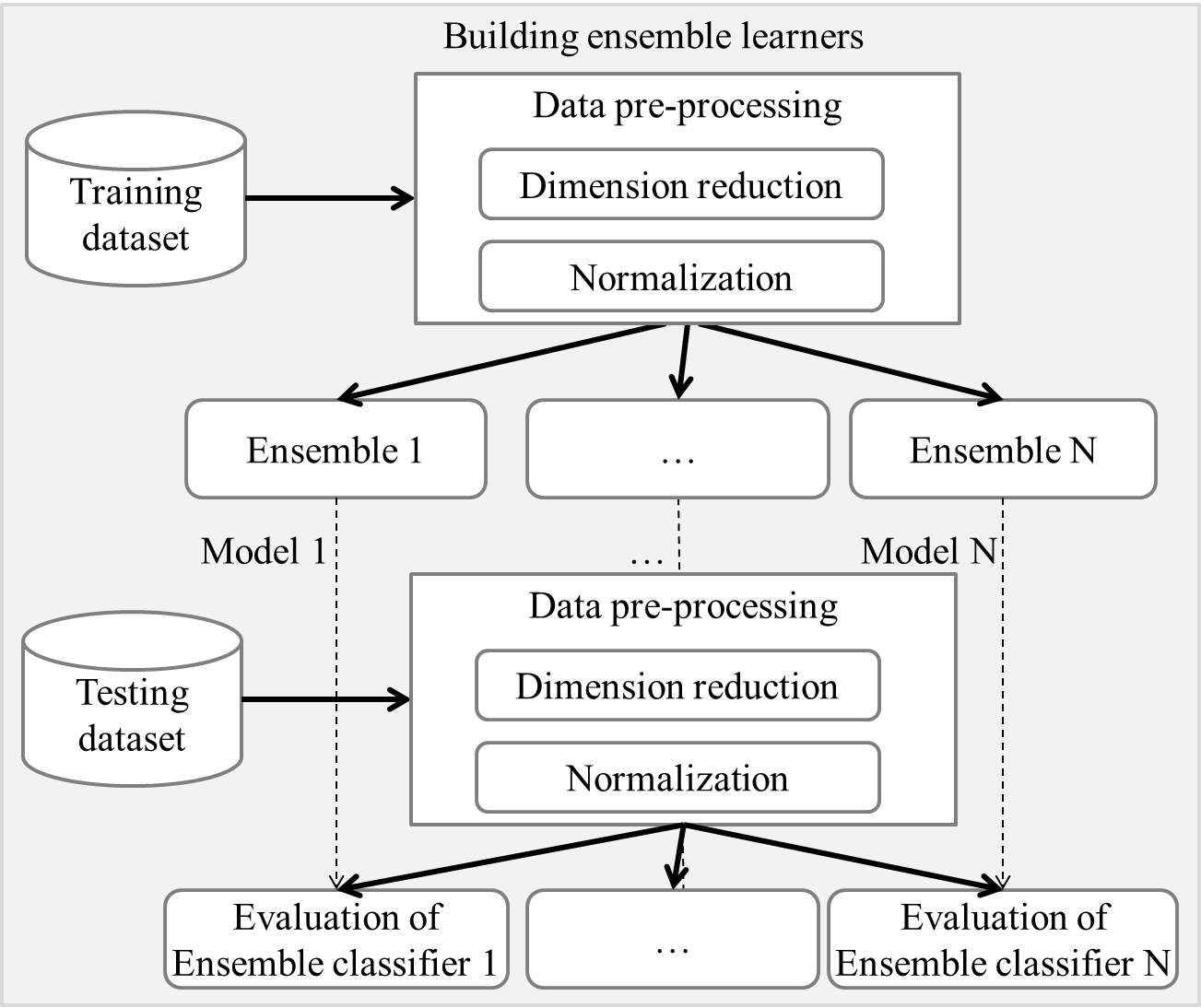}}
\caption{Building ensemble classifiers.}
\label{Building ensemble classifiers}
\end{figure}

Based on their diversities, we defined combinations of different learners to build multiexpert ensembles that employ voting functions. We also implemented learners using Random Forest classifier and bagging classifier. Furthermore, Multistage ensembles that use a boosting algorithm were also employed. For each ensemble method, we tuned some parameters (e.g., number of estimators) in order to have models with better classification accuracy. These ensemble methods have been evaluated and have resulted to significant accuracy improvement over the base learners. The obtained accuracy is also better than most of the solutions proposed in the literature. Fig.~\ref{Building ensemble classifiers} describes the creation of ensemble classifiers.

Combining machine learning methods leads to computationally intensive solutions. Therefore, as a second contribution, we proposed a new deployment architecture, detailed in the next subsection, in order to take advantage of the ensemble methods without largely increasing the system latency.

\subsection{The Deployment Architecture}
The more complex the combination of base learners is, the more computationally intensive the ensemble learner is. Thus, in a fog-to-things environment, we proposed to split the problem into two tasks and distribute it between the fog nodes and the cloud. Fog computing extends the cloud computing and has a suitable architecture to run applications that require both low latency and considerable computation resources  \cite{bonomi2012fog}. In the context of IoT, with the high number of latency-sensitive applications, fog computing could be an inevitable solution \cite{sarkar2018assessment}. The deployment of the IoT applications in a fog computing architecture is also known as fog-to-thing solutions.

Anomaly detection essentially requires very low latency in order to allow fast response for reducing the potential damage an anomalous traffic can cause. Since it is a binary classification, it will also require a less complex model than a multi-class classification. Thus, we built an anomaly detection model using the ensemble method discussed in the previous section. This model is deployed in fog nodes as a first level classifier. When an anomaly is detected, the security administrator receives an alert in order to take urgent and temporary measures. Then, the anomalous traffic is sent to the cloud for the second task.

The attack classification requires a more complex model, demanding more resources. Moreover, this task is less latency-sensitive than the first one. In this vein, we built an attack classification model, using also the ensemble method, that we deployed in the cloud as a second level classifier. When the attack category is predicted, the information is sent to the security administrator in order to apply complementary and more precise prevention mechanisms. 

\begin{figure}[htbp]
\centering{\includegraphics[width=0.9\columnwidth]{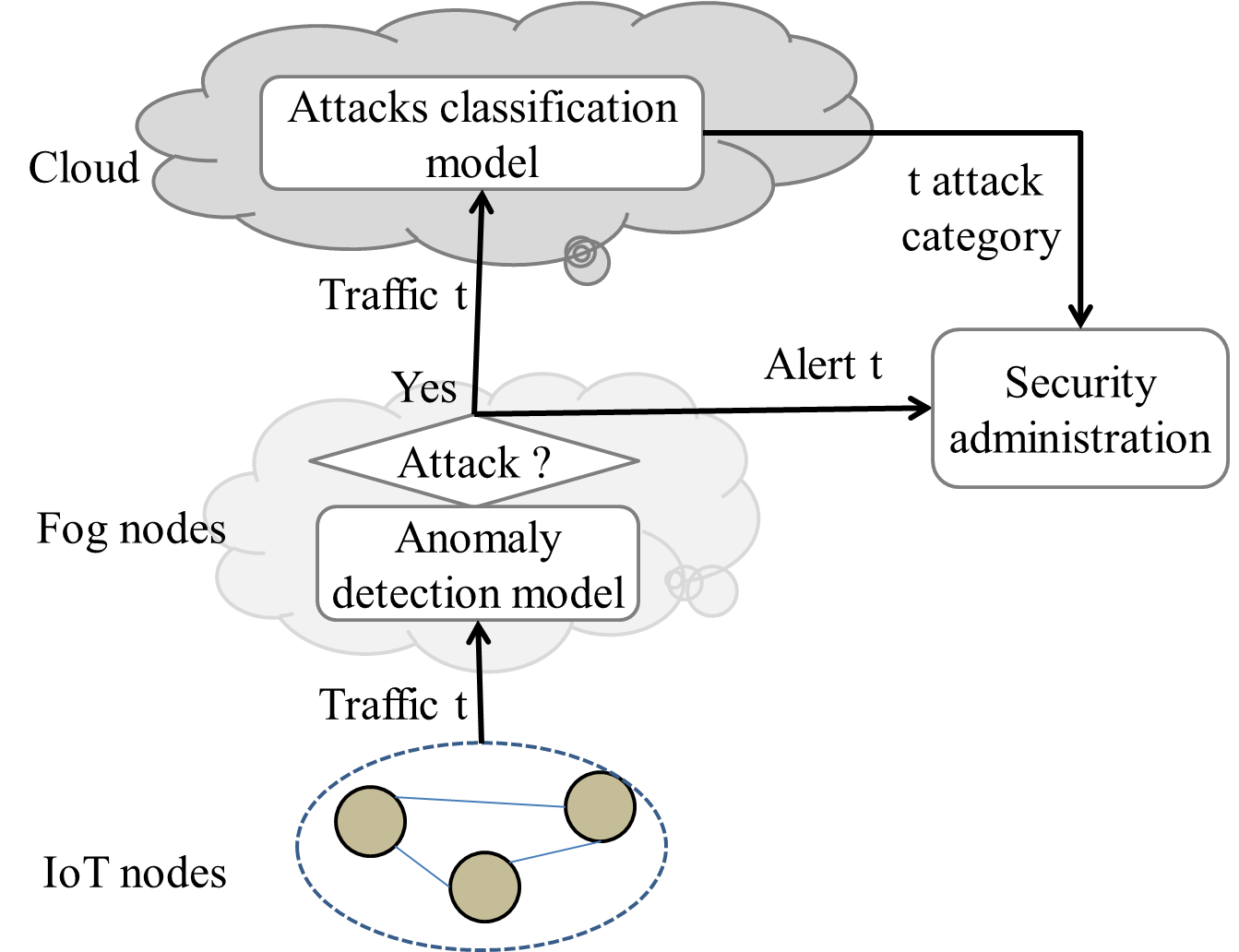}}
\caption{Deployment architecture of the proposed IDS in fog-to-things environment.}
\label{Our solution deployment in Fog-to-Things network}
\end{figure}

Our deployment architecture is illustrated in Fig.~\ref{Our solution deployment in Fog-to-Things network}. This architecture provides a faster intrusion detection system with attack classification for efficient prevention mechanisms. Our simulation results show that this new architecture is a promising solution for intrusion detection in a fog-to-things environment. The next two sections present the implementation of this approach using NSL-KDD dataset.
\vspace{-0.2 cm}
\section{Description of NSL-KDD dataset}
The NSL-KDD dataset is built from the KDDCUP\textquotesingle99 dataset that is also based on the data captured in DARPA\textquotesingle98 IDS evaluation program \cite{cup1999dataset}. KDDCUP\textquotesingle99 training dataset contains approximately 4,900,000 single connection vectors each of which contains 41 features and is labeled as either normal or an attack, with exactly one specific attack type. The specific attack types are regrouped into different attack categories, namely probing (Probe), Remote to Local (R2L), Denial of Service (DoS), and User to Root (U2R).
In the past, KDDCUP\textquotesingle99 dataset was widely used for evaluating machine learning models. However, one of its distinctive disadvantages was that even a simple machine learning model could easily have 99\% of accuracy without using any advanced tuning operation either at the pre-processing or the training phase. Under such conditions, it was quite difficult to compare different machine learning models on this dataset.

In 2009, the work in \cite{tavallaee2009detailed} conducted a statistical analysis on this dataset and found important issues. The first issue is the huge number of redundant records. About 78\% and 75\% of the records are duplicated in the train and test datasets, respectively. In train dataset, this caused learning algorithms to be biased towards the more frequent records, and thus prevent them from learning infrequent records. For the test dataset, the huge redundancies caused the evaluation results to be higher for the methods with better detection rates on the frequent records. The second problem is the low level of challenge.  After using 21 classifiers, most of the models gain a very high accuracy and very low false alarm rate. This was a problem in the way that it reduced the challenge and the advantage to compare different models. 
NSL-KDD is constituted by selection records of the complete KDD dataset in a way that avoids the aforementioned shortcomings. The first issue was resolved by eliminating all redundant records from the training and the testing datasets. To solve the second issue, NSL-KDD created two challenging test datasets, namely KDDTest+ and KDDTest-21. One increased just the challenge by reducing ``\textit{easy to classify records}'' (\textit{i.e.}, records successfully classified by most of the 21 models employed by the authors) and the second that eliminated all the records that were ``\textit{the easiest to classify}'' (\textit{i.e.}, records successfully classified by all the 21 models), making definitely difference between different models. Table \ref{NSL-KDD dataset} presents the NSL-KDD train dataset (KDDTrain+) and the two test datasets (KDDTest+ and KDDTest-21).

\begin{table}[htbp]
\caption{Organization of the NSL-KDD dataset.}
\begin{center}
{\tabulinesep=1.2mm
\begin{tabu}{|c|c|c|c|}
\hline
\textbf{}&\multicolumn{3}{|c|}{\textbf{Number of records used in training and testing datasets}} \\
\cline{2-4} 
\textbf{Class} & \textbf{\textit{KDDTrain+}}& \textbf{\textit{KDDTest+}}& \textbf{\textit{KDDTest-21}} \\
\hline
Normal & 67343 & 9711 & 2152 \\
\hline
DOS    & 45927    & 7458    & 4342\\
\hline
Probe    & 11656    & 2421    & 2402 \\
\hline
R2L    & 995    & 2754    & 533 \\
\hline
U2R    & 52    & 200    & 2421 \\
\hline
Total    & 125973    & 22544    & 11850 \\
\hline
\end{tabu}}
\label{NSL-KDD dataset}
\end{center}
\end{table}
\vspace{-0.2 cm}
\section{Experimentation results and discussions}
This section presents the experimentation carried out and the results obtained. It also discusses these results and provides a comparison with the previous works. The experimentation is done at two stages, \textit{i.e.}, the anomaly detection stage then attack classification stage. All the tasks are performed using the Python programming language and Scikit-learn (the free machine learning library). The computer used is equipped with Intel(R) Xeon(R) CPU E3-1225 v6 and 16.0GB RAM.

\subsection{The Anomaly detection stage}
We began the implementation with the anomaly detection stage. Table \ref{Organisation of the NSL-KDD datasets for the binary classification} presents the organization of the NSL-KDD dataset for the binary classification.

\begin{table}[htbp]
\caption{Organization of the NSL-KDD dataset for the binary classification.}
\begin{center}
{\tabulinesep=1.2mm
\begin{tabu}{|c|c|c|c|}
\hline
\textbf{}&\multicolumn{3}{|c|}{\textbf{Number of records used in training and testing datasets}} \\
\cline{2-4} 
\textbf{Class} & \textbf{\textit{KDDTrain+}}& \textbf{\textit{KDDTest+}}& \textbf{\textit{KDDTest-21}} \\
\hline
Normal & 67343 & 9711 & 2152 \\
\hline
Attack    & 58630    & 12833    & 9698 \\
\hline
Total    & 125973    & 22544    & 11850 \\
\hline
\end{tabu}}
\label{Organisation of the NSL-KDD datasets for the binary classification}
\end{center}
\end{table}

\begin{figure}[htbp]
    \includegraphics[width=\columnwidth]{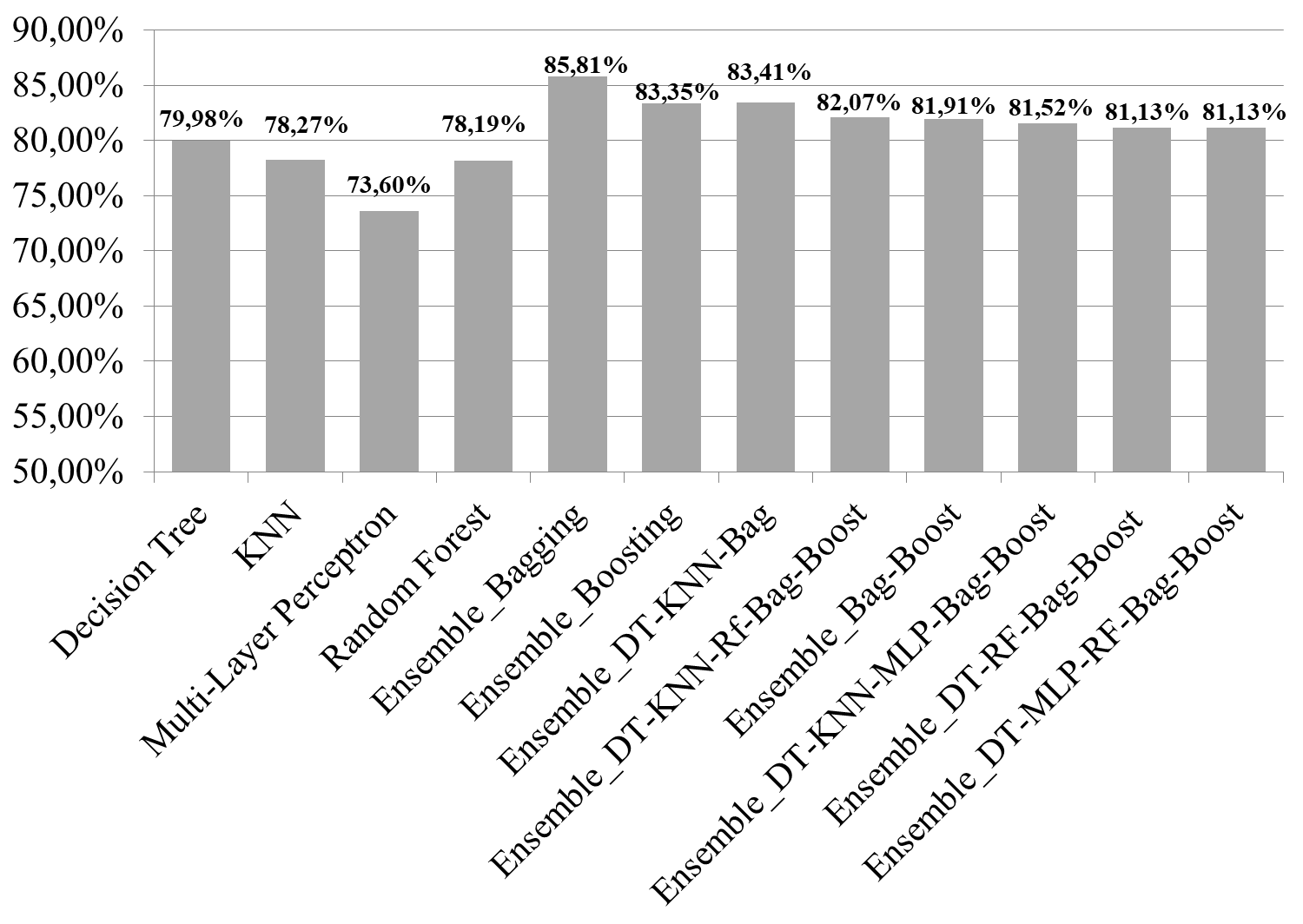}
    \caption{Accuracy of anomaly detection models using KDDTest+.}
\label{Accuracy of anomaly detection models using NSL-KDDTest+}
\end{figure}

\begin{figure}[htbp]
\centering{\includegraphics[width=\columnwidth]{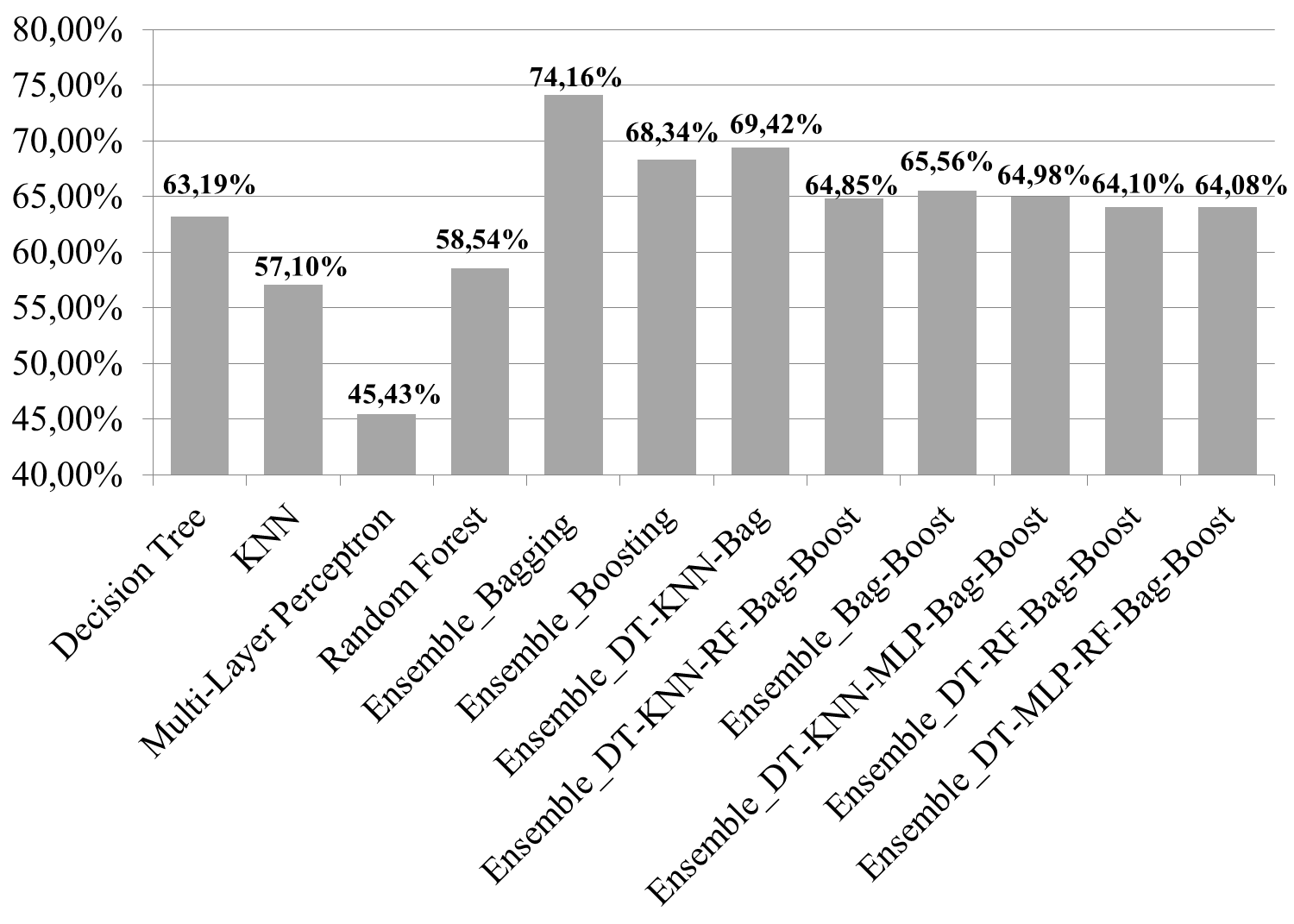}}
\caption{Accuracy of anomaly detection models using KDDTest-21.}
\label{Accuracy of anomaly detection models using NSL-KDDTest-21}
\end{figure}

Three base classifiers have been implemented including Decision Tree classifier (DT), K-Nearest Neighbour classifier (KNN), and Multi-Layer Perceptron classifier (MLP). Four ensemble methods have been implemented including Random Forest classifier (RF), Bagging Classifier, AdaBoost Classifier, and Voting Classifiers. In the Bagging and Boosting methods, Decision Tree classifiers are used as base learners. In the voting methods, we made different combinations of the three base learners and the three other ensembles according to the difference they presented in the prediction of each of the two classes.
All the models were trained using KDDTrain+ with 38 selected features, based on the work defined in \cite{bajaj2013improving}. As the data normalization did not provide a significant improvement on the accuracy of most of the models, we did not use it at the anomaly detection stage. For each model, we slightly tuned the hyper-parameters and kept the best ensembles. Advanced tuning methods were not employed as we want base learners to be just reasonably accurate, because they are not combined according to their individual accuracy, but for their diversity.
Fig. \ref{Accuracy of anomaly detection models using NSL-KDDTest+} gives the details of all models used and their accuracy with KDDTest+ dataset, then Fig. \ref{Accuracy of anomaly detection models using NSL-KDDTest-21} gives the accuracy of these models with KDDTest-21 dataset. For each model the accuracy showed is the one got with the tuned hyper-parameters.

The Bagging classifier provided the highest accuracy for anomaly detection. In our environment, the training time was less than 7 seconds and the prediction time for all the KDDTest+ records was less than 3 seconds.
Table \ref{Comparison of solutions using NSL-KDD dataset for binary classification} compares this result with previous works on anomaly detection using the same train and test datasets. As some works did not test their models with KDDTest-21, these cells are filled with the symbol ``-''. 
\begin{center}
\begin{table}[htbp]
\caption{Comparison of the accuracy of solutions using NSL-KDD dataset for binary classification.}
{\tabulinesep=1.2mm
\begin{tabu}{|l|c|c|} 
\hline
\textbf{Methods used}&\textbf{KDDTest+}&\textbf{KDDTest-21}\\
\hline
\begin{minipage}{4.5cm} Decision Tree Bagging Ensemble \\(Proposed method) \end{minipage} & 85.81 \% & 74.16 \%\\
\hline
    \begin{minipage}{4.5cm}
        Four base learners and ensembles \cite{paulauskas2017analysis}\newline Ensemble\_J48\_PART \newline PART \newline J48 \newline C5.0 \newline NB
    \end{minipage} &  
    \begin{minipage}{1cm}
        -\newline 84.84 \% \newline 82.61 \% \newline 82.13 \% \newline 77.94 \% \newline 77.26 \% 
    \end{minipage} & 
    - \\ 
 \hline
  \begin{minipage}{4.5cm} Recurrent neural networks \cite{yin2017deep} \end{minipage} & 83.28 \% & 68.55 \% \\ 
 \hline
  \begin{minipage}{4.5cm} ANN with tansig transfer function, Levenberg-Marquardt (LM) and BFGS quasi-Newton Backpropagation (BFG) algorithm \cite{ingre2015performance} \end{minipage} & 79.9 \% & - \\ 
 \hline
  \begin{minipage}{4.5cm}  Self-Organization  Map (SOM)  Artificial  Neural  Network \cite{ibrahim2013comparison} \end{minipage} &  75.49 \% & -\\ 
 \hline
\end{tabu}}
\label{Comparison of solutions using NSL-KDD dataset for binary classification}
\end{table}
\end{center}
First, these results confirm that ensemble models can provide better performance than using a single complex model. Secondly, we can see that anomaly detection does not require complex voting methods in order to have a good prediction model. So, a simple ensemble model can be used in the fog nodes for efficient anomaly detection. The attack classification task can be performed as a complementary task in the cloud.

\subsection{The Attack classification stage}
The second part of the experimentation concerned the attack classification.  Table \ref{Organisation of the NSL-KDD dataset for four categories classification} presents the organization of the NSL-KDD dataset for this four categories classification.

\begin{table}[htbp]
\caption{Organization of the NSL-KDD dataset for four categories classification.}
\begin{center}
{\tabulinesep=1.2mm
\begin{tabu}{|c|c|c|c|}
\hline
\textbf{}&\multicolumn{3}{|c|}{\textbf{Number of records used in training and testing datasets}} \\
\cline{2-4} 
\textbf{Class} & \textbf{\textit{KDDTrain+}}& \textbf{\textit{KDDTest+}}& \textbf{\textit{KDDTest-21}} \\
\hline
DOS    & 45927    & 7458    & 4342\\
\hline
Probe    & 11656    & 2421 & 2402 \\
\hline
R2L    & 995    & 2754    & 533 \\
\hline
U2R    & 52    & 200    & 2421 \\
\hline
Total    & 58630    & 12833    & 9698 \\
\hline
\end{tabu}}
\label{Organisation of the NSL-KDD dataset for four categories classification}
\end{center}
\end{table}

\begin{figure}[htbp]
\centering{\includegraphics[width=\columnwidth]{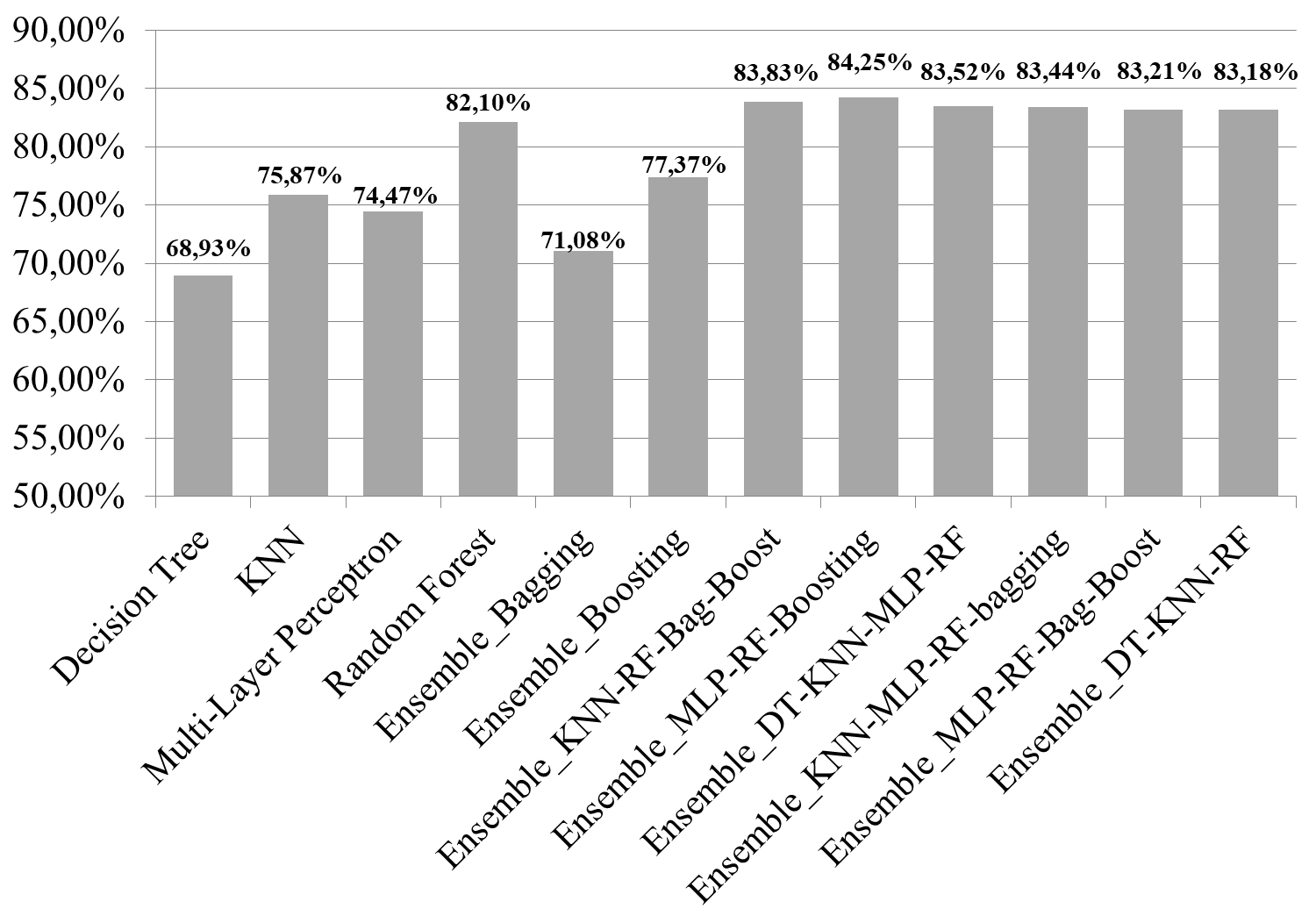}}
\caption{Accuracy of attack classification models using KDDTest+.}
\label{Accuracy of attacks classification models using NSL-KDDTest+}
\end{figure}

\begin{figure}[htbp]
\centering{\includegraphics[width=\columnwidth]{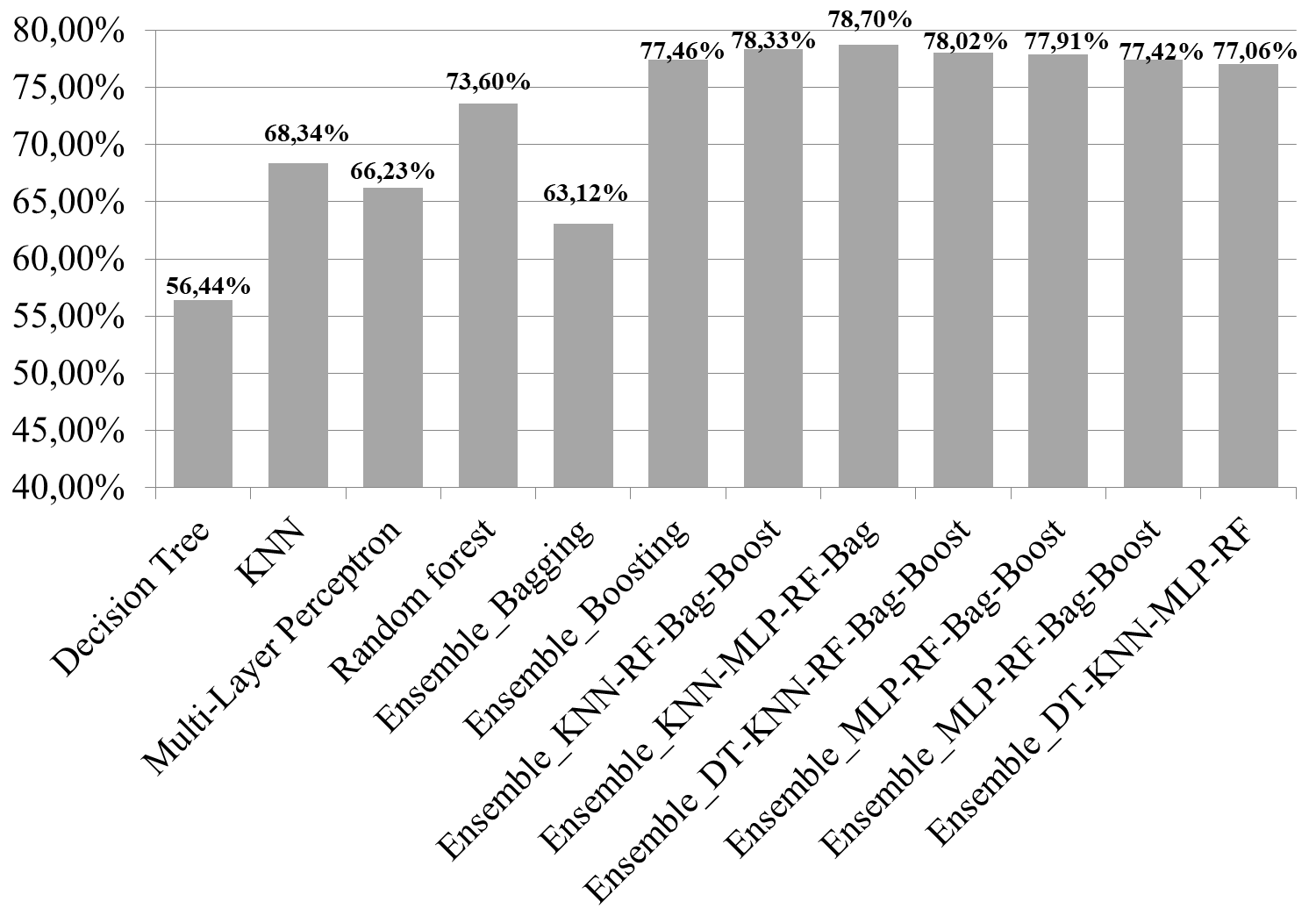}}
\caption{Accuracy of attack classification models using KDDTest-21.}
\label{Accuracy of attacks classification models using NSL-KDD Test-21}
\end{figure}

As mentioned in the anomaly detection stage, different methods have been implemented including base methods (Decision Tree, KNN, MLP) and ensemble methods (Random Forest, Bagging, Boosting, and Voting). All the models were trained using KDDTrain+ with 38 features selected. As the data normalization provided a significant improvement on the accuracy of most of the models, we used it for this stage. For each model, we slightly tuned the hyper-parameters and kept the best ensembles.
Fig. \ref{Accuracy of attacks classification models using NSL-KDDTest+} gives the details of all models used and their accuracy in KDDTest+ dataset then Fig. \ref{Accuracy of attacks classification models using NSL-KDD Test-21} gives the accuracy of these models on KDDTest-21 dataset. For each model the accuracy showed is the one got with the tuned hyper-parameters.

The voting ensemble using MLP, Random Forest, Bagging and Boosting as base learners gives the best accuracy on KDDTest+ dataset while the voting method using KNN, MLP, Random Forest and Bagging as base learners gives the best accuracy with KDDTest-21 dataset. However, the best model for the two NSL-KDD test datasets is the voting ensemble using KNN, Random Forest, Bagging and Boosting. This voting ensemble provides 83.83\% of accuracy and 78.33\% of accuracy in KDDTest+ and KDDTest-21, respectively. So, we proposed this model for the attack classification stage. 

Previous works that tried multi-class classification with NSL-KDD performed it using one model for five-categories classification (Probing, DoS, R2L, U2R, and Normal). Table \ref{Comparison of solutions using NSL-KDD dataset for multi-class classification} gives a comparison of these solutions with our attack classification model (Four-categories classification). As some works did not test their models with KDDTest-21, these cells are filled with the symbol ``-''. 

\begin{center}
\begin{table}[htbp]
\caption{Comparison of solutions using NSL-KDD dataset for multi-class classification.}
{\tabulinesep=1.2mm
\begin{tabu}{|l|c|c|} 
\hline
\textbf{Methods used}&\textbf{KDDTest+}&\textbf{KDDTest-21}\\
\hline
    \begin{minipage}{4.5cm} 
        Voting Ensemble using KNN, Random Forest, Bagging and Boosting of decision trees. Four-categories classification. (Proposed method).
    \end{minipage} & 
    83.83 \% & 78.33 \%\\
\hline
    \begin{minipage}{4.5cm}  
        Five methods \cite{tavallaee2009detailed} : \newline NBTree \newline Random Tree \newline Decision Trees J48 \newline Random Forest \newline Multi-Layer Perceptron
    \end{minipage} &  
    \begin{minipage}{1cm} 
        -\newline 82.02 \% \newline 81.59 \% \newline 81.05 \% \newline 80.67 \% \newline 77.41 \%
    \end{minipage} & 
    \begin{minipage}{1cm} 
        -\newline 66.16 \% \newline 63.97 \% \newline 63.26 \% \newline 58.51 \% \newline 57.34 \%
    \end{minipage}\\ \hline
    \begin{minipage}{4.5cm} 
        Six methods \cite{bajaj2013improving}: \newline SimpleCart \newline J48 \newline NBTree \newline Naïve Bayes \newline Multi-layer Perception \newline LibSVM

    \end{minipage} & 
    \begin{minipage}{1cm} 
    - \newline 82.32 \% \newline 81.93 \% \newline 80.67 \% 75.78 \% \newline 73.54 \% \newline 71.02 \%
    \end{minipage} &
    \begin{minipage}{1cm} 
        -\newline 66.77 \% \newline 65.65 \% \newline 63.62 \% \newline 54.25 \% \newline 56.28 \%\newline 44.84 \%
    \end{minipage} \\
\hline
\begin{minipage}{4.5cm} 
        Recurrent neural networks \cite{yin2017deep}.
    \end{minipage} & 
    81.29 \% & 64.67  \%\\
\hline
\begin{minipage}{4.5cm} 
        ANN with tansig transfer function, Levenberg-Marquardt (LM) and BFGS quasi-Newton Backpropagation \cite{ingre2015performance}.
    \end{minipage} & 
    81.20 \% &  -\\
\hline
\end{tabu}}
\label{Comparison of solutions using NSL-KDD dataset for multi-class classification}
\end{table}
\end{center}

For the best of our knowledge, our approach gives the best attack classification model on the KDDTest+ and NSL-KDD-21 datasets. The proposed model uses a complex voting ensemble but as this task will be done in the cloud, the complexity can be handled thanks to the availability of more computation resources. In our experimentation environment, the training took maximum 80 seconds and the prediction for the whole KDDTest+ records took maximum 15 seconds.

Our experimental results show that this architecture is an efficient approach for intrusion detection a in the fog-to-things context in order to have a detection with low latency for fast reaction and accurate attack classification for precise correction using the required prevention mechanisms according to the attack categories.
\vspace{-0.2 cm}
\section{Conclusion}
This work demonstrated that ensemble learners result to better classification models for the NSL-KDD dataset, the most realistic and challenging dataset available for machine learning based intrusion detection. The proposed solution employed diverse base learners using different known algorithms and built different ensemble classifiers for anomaly detection and attack classification. Overall accuracy of 85.81\% and 84.25\% have been achieved for the binary classification and attack classification, respectively. A deployment architecture was also proposed where anomaly detection was performed in fog nodes to allow faster detection and response, and attack classification in the cloud to benefit from more resources to run a more complex ensemble for better classification that guides the intrusion prevention tasks. Investigating more diverse base learners and various combination methods can improve better these results and is envisioned as future work. 

\bibliography{bibliography} 
\bibliographystyle{ieeetr}

\end{document}